\documentclass[prl,reprint,twocolumn,superscriptaddress,floatfix,showpacs,10pt]{revtex4-1}%
\usepackage{amsfonts,amsmath,amssymb}
\usepackage{slashed}
\usepackage[colorlinks=true,urlcolor=blue,anchorcolor=blue,citecolor=blue,filecolor=blue,linkcolor=blue,menucolor=blue,pagecolor=blue,linktocpage=true,pdfproducer=medialab,pdfa=true]{hyperref}

\begin{document}

\title{Poisson-Lie T-Duality in Double Field Theory}

\newcommand{\be}{\begin{equation}}
\newcommand{\ee}{\end{equation}}
\newcommand{\ben}{\begin{displaymath}}
\newcommand{\een}{\end{displaymath}}
\newcommand{\bea}{\begin{eqnarray}}
\newcommand{\eea}{\end{eqnarray}}
\newcommand{\nn}{\nonumber}
\newcommand{\non}{\nonumber\\}
\newcommand{\bean}{\begin{eqnarray*}}
\newcommand{\eean}{\end{eqnarray*}}
\newcommand{\beqs}{\begin{eqnarray}}
\newcommand{\eeqs}{\end{eqnarray}}

\newcommand{\nin}[1] {\underline{\phantom{h}}\hskip-6pt {#1}}

\author{Falk Hassler}
\email{fhassler@unc.edu}
\affiliation{University of North Carolina,
Department of Physics and Astronomy,
Chapel Hill, NC 27599-3255, USA}
\affiliation{University of Pennsylvania,
Department of Physics and Astronomy,
Philadelphia, PA 19104-6396, USA}

\begin{abstract}
We present a formulation of Double Field Theory with a Drinfeld double as extended spacetime. It makes Poisson-Lie T-duality (including abelian and non-abelian T-duality as special cases) manifest. This extends the scope of possible applications of the theory, which so far captured abelian T-duality only, considerably. The full massless bosonic subsector (NS/NS and R/R) of type II string theories is covered.
\end{abstract}

\maketitle

Superstring theory is a very successful framework to explore quantum field theories and quantum gravity. Dualities lie at its heart by connecting the five perturbative formulations and M-theory. Especially T-duality is inherently linked to the extended nature of the string. Unfortunately, this term is a bit ambiguous because it is used in different contexts. Normally, it refers to abelian T-duality \cite{Buscher:1987sk} which requires the two target spaces it links to have abelian isometries. But there is also a non-abelian counterpart (NATD) \cite{delaOssa:1992vci,*Giveon:1993ai,*Alvarez:1994np}. It only holds at the classical level but still admits to construct new string backgrounds, for example \cite{Lozano:2012au,*Itsios:2013wd}. This provides an important tool in studying the underlying structure of the AdS/CFT correspondence. Like dualities are essential for string theory, integrability plays a similar distinguished role in this endeavor. Considering the close relation between string theory and AdS/CFT, it is not surprising that non-abelian T-duality also gives access to deformations of integrable models \cite{Hoare:2016wsk,*Borsato:2016pas}. Based on these three pillars, NATD, AdS/CFT and integrability, many interesting results have been presented recently.

T-duality is however not directly manifest in the low-energy effective target space description, 10D supergravity (SUGRA). Both variants, abelian and non-abelian, are meditated by involved, non-linear transformations of the fields. Double Field Theory (DFT) \cite{Hull:2009mi,*Hull:2009zb, Hohm:2010jy,*Hohm:2010pp} (see also \cite{Tseytlin:1990va,*Siegel:1993th}) is an approach to make abelian T-duality manifest by formally doubling the complete target space. Physically, this doubling is only relevant for directions with capture the abelian isometries of the background. They form a flat $d$-torus, while the additional coordinates are dual to string modes winding along non-contractible cycles of this torus. Abelian T-duality exchanges momentum with winding modes and therefore is a simple O($d$,$d$) rotation of the coordinates in the doubled space. For a background independent, consistent formulation of the theory \cite{Hohm:2010jy} the section condition has to be imposed. It guarantees that the theory can be reduced to the common SUGRA description (at least in one duality frame). DFT has become an indispensable tool for studying consistent truncations, non-geometric backgrounds, $\alpha'$ correction and many other questions related to abelian T-duality.

Nevertheless, it does not help with NATD and thus did not enter the line of research outlined above (a notable exception is \cite{Sakamoto:2017cpu}). Hence, objective of this letter is to present a version of DFT with manifest abelian and non-abelian T-duality. In fact it implements the even more general Poisson-Lie T-duality \cite{Klimcik:1995ux,Klimcik:1995jn} which includes the former as special cases. We discuss the full bosonic part of type IIA/B SUGRA, also including the R/R sector.

\paragraph*{Poisson-Lie T-duality:}
As it is the starting point for all following derivations, let us review the salient features of Poisson-Lie T-duality. We start from a bosonic $\sigma$-model with target space $M$ which admits a free action of the Lie-group $G$. It is governed by the action
\begin{equation}\label{eqn:sigmamodel}
  S = \int d z d \bar z \, E_{ij} \partial x^i \bar\partial x^j
\end{equation}
where $E_{ij}=g_{ij} + B_{ij}$ combines the metric on $M$ with the anti-symmetric $B$-field. If we vary this action with respect to infinitesimal changes of the target space coordinates $\delta x^i = v_a{}^i  \delta\epsilon^a$, where $v_a{}^i$ is the left-invariant vector field on $G$, we find
\begin{equation}\label{eqn:varS}
  \delta S = \int d z d \bar z \,  L_{v_a} E_{ij} \, \partial x^i \bar\partial x^j \, \delta\epsilon^a - \int d J_a \, \delta\epsilon^a
\end{equation}
with the Noether currents
\begin{equation}\label{eqn:Ja}
  J_a = - v_a{}^i E_{ji} \partial x^j d z + v_a{}^i E_{ij} \bar\partial x^j d\bar z\,.
\end{equation}
In order to permit Poisson-Lie T-duality, these currents have to satisfy the on-shell integrability condition
\begin{equation}\label{eqn:intJa}
  d J_a - \frac12 F^{bc}{}_a J_b \wedge J_c = 0
\end{equation}
where $F^{bc}{}_a$ denotes the structure constants of the Lie-algebra $\tilde{\mathfrak{g}}$ corresponding to the dual Lie-group $\tilde G$. On-shell, the variation \eqref{eqn:varS} vanishes. In combination with the integrability condition for $J_a$, this gives rise to the constraint
\begin{equation}\label{eqn:LvaEij}
  L_{v_a} E_{ij} = - F^{bc}{}_a v_b{}^k v_c{}^l E_{ik} E_{lj}
\end{equation}
on the background. This equations reveals a remarkable feature of Poisson-Lie T-duality: In contrast to its abelian/non-abelian descendants, it does not require isometries. Of course, it still works with them present. In this case, the dual Lie-group is abelian and $d J_a$=0 holds. But in general the $\sigma$-model admits a non-commutative conservation law \cite{Klimcik:1995ux}.

As $v_a{}^i$ is the left-invariant vector field on $G$, its Lie derivative generates the corresponding Lie-algebra $[L_{v_a}, L_{v_b}] = L_{F_{ba}{}^c v_c}$, denoted as $\mathfrak{g}$,  with structure constants $F_{ab}{}^c$. By combining \eqref{eqn:LvaEij} with this commutator, we obtain a compatibility relation between the structure constants of $\mathfrak{g}$ and $\tilde{\mathfrak{g}}$. It requires that these two groups are the maximally isotropic subgroups of a Drinfeld double $\mathcal{D}$ \cite{Drinfeld:1986in} with the bilinear, invariant pairing
\begin{equation}\label{eqn:invpairing}
  \langle t_a , t^b \rangle = \delta_a^b \,, \quad
  \langle t^a , t_b \rangle = \delta^a_b \,, \quad
  \langle t_a , t_b \rangle = \langle t_b , t_a \rangle = 0
\end{equation}
where $t_a$ and $t^a$ are the generators of $G$ and $\tilde G$, respectively. This double is the central object behind the whole construction. The dual $\sigma$-model
\begin{equation}\label{eqn:sigmamodeldual}
  \tilde S = \int dz d\bar z \, \tilde E_{ij} \partial x'^i \bar\partial x'^j
\end{equation}
arises after exchanging its two subgroups $G$ and $\tilde G$. Due to the integrability condition \eqref{eqn:intJa}, solutions $g(z,\bar z)\in G$ of the field equations for \eqref{eqn:sigmamodel} can be lifted to the Drinfeld double $\mathcal{D} \ni d(z,\bar z) = g(z,\bar z) \tilde g(z,\bar z)$ by
\begin{equation}
  J_a = d \tilde g(z,\bar z) \tilde g^{-1}(z, \bar z)\,.
\end{equation}
Using the alternative decomposition $d=\tilde h(z,\bar z) h(z,\bar z)$ provides a solution $\tilde h(z,\bar z)\in\tilde G$ for the dual model \cite{Klimcik:1995ux}. After changing from the Lagrangian description to the Hamiltonian, the two models are connected by a canonical transformation on their phase space variables \cite{Klimcik:1995dy,*Sfetsos:1997pi}. So at the classical level their dynamics is indistinguishable.

It is convenient to exploit the underlying structure of the Drinfeld double to find an explicit expression for the metric and the $B$-field by integrating \eqref{eqn:LvaEij}. At the unit element of the double, and therewith also of $G$, we use a constant, invertable $E^{0\,ab}$ as initial conditions. What is left to do, is to transport it by the adjoint action to all other points $g\in G$ \cite{Klimcik:1995ux}. More explicitly, if we denote the adjoint action as
\begin{equation}\label{gen:adjg}
  g \,t_a\, g^{-1} = M_a{}^b t_b \,, \quad
  g \,t^a\, g^{-1} = M^{ab} t_b + M^a{}_b t^b\,,
\end{equation}
one obtains%check against (1) of hep-th/9803019, using (4),(5),(6)
\begin{equation}\label{eqn:Eij}
  E^{ij} = v_c{}^i M_a{}^c (M^{ae} M^b{}_e + E^{0\,ab}) M_b{}^d v_d{}^j
\end{equation}
where $E^{ij}$ is the inverse of $E_{ij}$. For the dual model, the same procedure applies with the adjoint action of $\tilde g$ instead of $g$ and a map $E^0_{ab}: \tilde g \rightarrow g$ (the inverse of $E^{0\,ab}$) resulting in
\begin{equation}\label{eqn:tildeEij}
  \tilde E^{ij} = \tilde v^{c i} \tilde M^a{}_c ( \tilde M_{ae} \tilde M_b{}^e + E^0_{ab} ) \tilde M^b{}_d \tilde v^{d j}\,.
\end{equation}

A decomposition of the Drinfeld double in two maximally isotropic subgroups refers to a Manin triple. We are interested in the set $M(\mathcal{D})$ of these triples for a given double $\mathcal{D}$. It plays the role of the moduli space of dual $\sigma$-models and has at least two points originating from $G$/$\tilde G$ and swapping them \cite{Klimcik:1995jn}. Abelian and non-abelian T-duality are special cases of this construction. While the former arises if the double is abelian and has $M(\mathcal{D}) = \text{O($D$,$D$,$\mathbb{Z}$)}$, where $D$ denotes the dimension of the target space $M$, the latter results from a non-abelian $G$ and an abelian $\tilde G$.

\paragraph*{Double Field Theory on Drinfeld doubles:}\label{sec:DFTwzw}
So far, we approached Poisson-Lie T-duality from the world-sheet perspective. Main objective of this letter is to show how this symmetry can be made manifest in the corresponding target space low-energy effective theory. As the Drinfeld double is essential to construct all dual backgrounds, it should be a fundamental part of this theory. Remember that $\mathcal{D}$ is a Lie-group equipped with a bi-invariant linear form \eqref{eqn:invpairing} of split signature. Hence, the framework of DFT on group manifolds naturally applies. It was originally derived via Closed String Field Theory for a Wess-Zumino-Witten model \cite{Blumenhagen:2014gva} and afterward generalized to arbitrary, $2D$-dimensional group manifolds \cite{Blumenhagen:2015zma,*Bosque:2015jda} which admit an embedded into O($D$,$D$). Its action 
\begin{align}
  S_{\mathrm{NS}} =& \int d^{2D} X e^{-2d} \Big(  \frac{1}{8} \mathcal{H}^{CD} \nabla_C \mathcal{H}_{AB} \nabla_D \mathcal{H}^{AB} \nonumber \\
    & -\frac{1}{2} \mathcal{H}^{AB} \nabla_{B} \mathcal{H}^{CD} \nabla_D \mathcal{H}_{AC} - 2 \nabla_A d \nabla_B \mathcal{H}^{AB} \nonumber\\
    & + 4 \mathcal{H}^{AB} \nabla_A d \nabla_B d + \frac{1}{6} F_{ACD} F_B{}^{CD} \mathcal{H}^{AB} \Big)
    \label{eqn:Sdftwzw}
\end{align}
governs the dynamics of the generalized metric $\mathcal{H}^{AB}$, a symmetric, O($D$,$D$) valued matrix, and the generalized dilation $d$ on the extended space $\mathcal{D}$. The indices $A$ to $F$ label the adjoint representation of the corresponding Lie-algebra $\mathfrak{d}$. They are lowered/raised with the O($D$,$D$) invariant metric $\eta_{AB}$ and it inverse, which directly results from the pairing \eqref{eqn:invpairing} as $\langle t_A, t_B \rangle = \eta_{AB}$. Furthermore, we need the $\eta$-compatible, covariant derivative
\begin{equation}\label{eqn:covderiv}
  \nabla_A V^B = D_A V^B + \frac13 F_{AC}{}^B V^C - w F_A V^B \,,
\end{equation}
here for a vector density $V^A=( V_a \,\, V^a )$ of weight $w$, in order to evaluate the action. It incorporates the structure constants $F_{AB}{}^C$ of $\mathfrak{d}$ and the flat derivative
\begin{equation}
  D_A = E_A{}^I \partial_I\,, \quad \text{with} \quad [D_A, D_B] = F_{AB}{}^C D_C\,,
\end{equation}
expressed in terms of a partial derivative and the right-invariant vector field $E_A{}^I$ on $\mathcal{D}$ whose coordinates are $X^I=(\tilde x_i \,\, x^i)$. Furthermore, there is the density part with $F_A = D_A \log E$ and $E=\det E^A{}_I$. The action \eqref{eqn:Sdftwzw} is invariant under a local O($D$,$D$) symmetry mediated by the generalized Lie derivative
\begin{equation}\label{eqn:genLie}
  \mathcal{L}_\xi V^A = \xi^B \nabla_B V^A + \big( \nabla^A \xi_B - \nabla_B \xi^A \big) V^B + w \nabla_B \xi^B V^A
\end{equation}
which extends in the usual way to higher rank tensors like $\mathcal{H}^{AB}$ ($w=0$), and $e^{-2 d}$ ($w=1$). Closure of these infinitesimal transformations requires to impose the section condition
\begin{equation}\label{eqn:SC}
  D_A \, \cdot \, D^A \, \cdot = 0
\end{equation}
on arbitrary combinations of physical fields and gauge parameters, formally denoted by $\cdot$\,.

A technique to solve \eqref{eqn:SC} was introduced in \cite{Hassler:2016srl}. It selects $D$ physical coordinates $x^i$ on the group manifold using a maximally isotropic subgroup. If we apply it to a Drinfeld double $\mathcal{D}$, group elements are parameterized by $d = g(x^i) \tilde g(\tilde x_i)$ and the target space $M$ is identified with the coset $\mathcal{D}/\tilde G$ which is nothing else than $G$. All physical fields and parameters of gauge transformations are restricted to depend on $x^i$ only. Thus, the section condition \eqref{eqn:SC} is solved. A refinement of this statement is required for tensor densities, like the generalized dilaton. They split into a background and a fluctuation part. Here the section condition \eqref{eqn:SC} only applies for the latter \cite{Blumenhagen:2015zma}.

Finally, we need to connect the generalized metric $\mathcal{H}^{AB}$ and the GL($D$) element $E_{ij}$ used in the $\sigma$-model \eqref{eqn:sigmamodel}. Latter decomposes into the metric $g_{ij}$ and the $B$-field $B_{ij}$ which are embedded in the generalized metric 
\begin{equation}\label{eqn:genmetric}
  \widehat{\mathcal{H}}^{\hat I\hat J} = \begin{pmatrix} g_{ij} - B_{ik} g^{kl} B_{lk} & -B_{ik} g^{kl} \\
    g^{ik} B_{kj} & g^{ij}
    \end{pmatrix}
\end{equation}
on the generalized tangent bundle $T M \oplus T^* M$ over $M$. In order to clearly distinguish this bundle from $T \mathcal{D}$, the corresponding indices and quantities are decorated with a hat. $\widehat{\mathcal{H}}^{\hat I\hat J}$ is linked to it flat counterpart by the generalized frame field
\begin{equation}\label{eqn:genframefield}
  \widehat{E}_A{}^{\hat I} = M_A{}^B \begin{pmatrix} v^b{}_i & 0 \\
    0 & v_b{}^i \end{pmatrix}{}_B{}^{\hat I} \,, 
  \quad  M_A{}^B t_B = g \, t_A \, g^{-1} \,,
\end{equation}
as $\widehat{\mathcal{H}}^{\hat I\hat J} = \widehat{E}_A{}^{\hat I} \mathcal{H}^{AB} \widehat{E}_B{}^{\hat J}$. Note that the components of $M_A{}^B$ were already introduced in \eqref{gen:adjg}. Taking furthermore $E^0_{ab} = g_{ab} + B_{ab}$ and writing $\mathcal{H}^{AB}$ like \eqref{eqn:genmetric} but now with indices $a$, $b$ instead of $i$, $j$, we indeed exactly reproduce the background \eqref{eqn:Eij} derived from the $\sigma$-model.

It is instructive to rewrite the action \eqref{eqn:Sdftwzw} and it gauge transformations \eqref{eqn:genLie} in terms of these hatted quantities. For the generalized dilaton $d$, we also have to take into account the volume form on $\tilde{G}$ to introduce
\begin{equation}
  \widehat{d} = d + \frac12 \log \tilde v \,, \quad \tilde v = \det ( \tilde v^a{}_i )\,.
\end{equation}
A crucial part in the rewriting is to transform the covariant derivatives. It is straightforward to show that
\begin{align}\label{eqn:nablaflattogentang}
  \nabla_A V^B &\rightarrow \partial_{\hat I} \widehat{V}^{\hat J} + (\widehat{\Omega}_{[\hat I\hat K\hat L]} - \widehat{\Omega}_{\hat I\hat K\hat L}) \eta^{\hat L\hat J} \widehat{V}^{\hat K}\\
  \nabla_A d &\rightarrow \partial_{\hat I} \widehat{d} + \frac12 \widehat{\Omega}^{\hat J}{}_{\hat J\hat I}
    \quad \text{with} \quad
  \widehat{\Omega}_{\hat I\hat J\hat K} = \partial_{\hat I} \widehat{E}_{A\hat K} \widehat{E}^A{}_{\hat J}
    \nonumber
\end{align}
holds if we assume that both Lie-groups $G$ and $\tilde G$ are unimodular. As result we get the action
\begin{align}\label{eqn:SdftNS}
  S_{\mathrm{NS}} &= V_{\tilde G} \int d^D x e^{-2 \widehat{d}} \Big(  \frac{1}{8} \widehat{\mathcal{H}}^{\hat K\hat L} \partial_{\hat K} \widehat{\mathcal{H}}_{\hat I\hat J} \partial_{\hat L} \widehat{\mathcal{H}}^{\hat I\hat J} - \\
    & 2 \partial_{\hat I} \widehat{d} \partial_{\hat J} \widehat{\mathcal{H}}^{\hat I\hat J} -\frac{1}{2} \widehat{\mathcal{H}}^{\hat I\hat J} \partial_{\hat J} \widehat{\mathcal{H}}^{\hat K \hat L} \partial_{\hat L} \widehat{\mathcal{H}}_{\hat I\hat K}  + 
    4 \widehat{\mathcal{H}}^{\hat I\hat J} \partial_{\hat I} \widehat{d} \partial_{\hat J} \widehat{d} \Big)
    \nonumber
\end{align}
of conventional DFT and the corresponding generalized Lie derivative after neglecting boundary terms and taking into account the section condition $\partial_{\hat I} \,\cdot \, \partial^{\hat I} \, \cdot$\,. Evaluating the integral in \eqref{eqn:Sdftwzw} over the unphysical directions $\tilde x_i$ contributes the volume factor $V_{\tilde G}$. After suppressing it, as it is common practice in DFT, and following \cite{Hohm:2010jy}, we find the action
\begin{equation}
  S_\mathrm{NS} = \int\mathrm{d}^{D}x\,
    \sqrt{g} e^{-2\phi} \big(\mathcal R + 4 \partial_i \phi \partial^i \phi 
    - \frac{1}{12} H_{ijk} H^{ijk} \big)
\end{equation}
for the NS/NS sector of type II SUGRA with the curvature scalar $R$, the dilaton $\phi = \widehat{d} - 1/4 \log (g)$ and the three-form flux $H_{ijk} = 3 \partial_{[i} B_{jk]}$. Finally note that $\widehat{E}_A{}^{\hat I}$ is a globally defined, O($D$,$D$) valued generalized frame field which renders $M$ a generalized parallelizable space. This property is essential to obtain \eqref{eqn:SdftNS}, because it gives rise to $F_{ABC}=3 \widehat{\Omega}_{[ABC]}$.

For the R/R sector the action on $\mathcal{D}$,
\begin{equation}\label{eqn:SdftwzwRR}
  S_{\mathrm{R}} = \frac14 \int d^{2D} X\, (\slashed{\nabla} \chi)^\dagger \, S_{\mathcal{H}} \, \slashed{\nabla} \chi\,,
\end{equation}
is a straightforward generalization of \cite{Hohm:2011zr}. It incorporates the O($D$,$D$) Majorana-Weyl spinor $\chi$ and the spinor version $S_{\mathcal{H}}$ of the generalized metric. We exclude time-like T-dualities, so this map is well-defined. Our $\Gamma$-matrices satisfy the canonical anti-commutator relation $\{\Gamma^A,\Gamma^B\} = 2 \eta^{AB}$. Additionally, the spin connection of the covariant derivative has to be fixed. We do so by requiring $\nabla_A \Gamma^B = 0$, resulting in
\begin{equation}
  \nabla_A \chi = D_A \chi - \frac{1}{12} F_{ABC} \Gamma^{BC} \chi - \frac12 F_A  \chi
\end{equation}
under the assumption $D_A \Gamma^B =0$ and taking into account that $\chi$ is a density of weight 1/2. Generalized diffeomorphisms leave \eqref{eqn:SdftwzwRR} invariant and are mediated by the generalized Lie derivative
\begin{equation}
  \mathcal{L}_\xi = \xi^A \nabla_A \chi + \frac12 \nabla_A \xi_B \Gamma^{AB} \chi + \frac12 \nabla_A \xi^A \chi\,.
\end{equation}
Depending on its chirality, $\chi$ either captures the type IIA or IIB R/R gauge potentials. A convenient way to parameterize it is in terms of $p$-forms $C^{(p)}$ with even or odd degree as
\begin{equation}\label{eqn:formstospinor}
  \chi = \sum\limits_{p=0}^D \frac1{2^{p/2} \, p!} C^{(p)}_{a_1\dots a_p} \Gamma^{a_1} \dots \Gamma^{a_p} | 0 \rangle\,.
\end{equation}
Moreover, one has to impose the self-duality constraint $\slashed{\nabla} \chi = - C^{-1} S_{\mathcal{H}} \slashed{\nabla}\chi$ \cite{Hohm:2011zr}, where $C$ is the charge conjugation matrix. As for the generalized metric in the NS/NS sector, we finally employ the spinor version $S_{\widehat{E}}$ of the generalized frame field \eqref{eqn:genframefield} to make contact with the conventional DFT R/R action
\begin{equation}\label{eqn:SdftRR}
  S_\mathrm{R} = V_{\tilde G} \int d^{D} x \frac14 (\slashed{\partial} \widehat{\chi})^\dagger \, S_{\widehat{\mathcal{H}}} \, \slashed{\partial} \widehat{\chi}
\end{equation}
where $\widehat{\chi} = \tilde v^{-1/2} S_{\widehat{E}} \chi$. Again the transformation of the covariant derivatives \eqref{eqn:nablaflattogentang} plays the central role in this calculation. Here, it has to be adapted to spinors. After some $\Gamma$-matrix algebra ones find the simple identification
\begin{equation}
  \slashed{\nabla} \chi \rightarrow \sqrt{\tilde v} S^{-1}_{\widehat{E}} \, \slashed{\partial} \widehat{\chi}
\end{equation}%checked! Requires identity S_{\hat E} D_A S^{-1}_{\hat E} = 1/4 \Omega_{A\hat J\hat K} \hat\Gamma^{\hat J\hat K}
which makes the identity between \eqref{eqn:SdftwzwRR} and \eqref{eqn:SdftRR} manifest. Note that $\slashed{\partial}$ denotes the contraction with the $\Gamma$-matrices $\widehat{\Gamma}^{\hat I}$, while $\slashed{\nabla}$ takes $\Gamma^A$. They are related by $\widehat{\Gamma}^{\hat I} = \widehat{E}_A{}^{\hat I} S_{\widehat{E}} \Gamma^A S_{\widehat{E}}^{-1}$\,. Assuming that the section condition holds, which is the case if $\widehat{\chi}$ depends on the physical coordinates only, \eqref{eqn:SdftRR} reduces to the democratic R/R action of type IIA/B SUGRA \cite{Hohm:2011zr}.

\paragraph*{Dual backgrounds:}
At least in the NS/NS sector, all essential parts of DFT on $\mathcal{D}$ have a natural analog on the world-sheet. This connection extends to the dual background which arises after swapping $G$ and $\tilde G$. We require that each duality frame (with coordinates $X^I$ and $X'^I$) parameterizes the same group elements \begin{equation}\label{eqn:2Ddiff}
  \mathcal{D} \ni d = g(x^i) \tilde g(\tilde x_i ) = \tilde g(x'^i) g(\tilde x'_i)
\end{equation}
in the Drinfeld double. This equation describes the $2D$-diffeomorphism $X^I \rightarrow X'^I$ which is a manifest symmetry of our formulation \cite{Blumenhagen:2015zma}. It does not exists in conventional DFT. Requiring that in both duality frames $\mathcal{H}^{AB}$ and the fluctuations of $\chi$ and $d$ depend on the physical coordinates only, we have to choose them constant. This is in agreement with the world-sheet perspective, where $E^0_{ab}$ is constant, too. Furthermore, we adapt the generalized frame field
\begin{equation}\label{eqn:genframefielddual}
  \widetilde{\widehat{E}}_A{}^{\hat I} = \tilde M_{A B}
    \begin{pmatrix} \tilde v_{b i} & 0 \\
      0 & \tilde v^{b i} \end{pmatrix}{}^{B I}\,, \quad
    \tilde M_A{}^B t_B = \tilde g \, t_A \, \tilde g^{-1}\,,
\end{equation}
because $\tilde E_A{}^{\hat I}$ would depend on unphysical coordinates $\tilde x'_i$ after applying \eqref{eqn:2Ddiff}. However, \eqref{eqn:genframefielddual} still allows to reduce the action to \eqref{eqn:SdftNS} and \eqref{eqn:SdftRR}. Now, we find
\begin{equation}\label{eqn:trafogenmetric}
  \widetilde{\widehat{\mathcal{H}}}{}^{\hat I\hat J} = \widehat{O}^{\hat I}{}_{\hat M} \widehat{\mathcal{H}}^{\hat M\hat N} \widehat{O}^{\hat J}{}_{\hat N}
    \quad \text{with} \quad
  \widehat{O}^{\hat I}{}_{\hat J} = \widetilde{\widehat{E}}_A{}^{\hat I} \widehat{E}^B{}_{\hat J}
\end{equation}
for the generalized metric which specifies the metric $\tilde g_{ij}$ and the $B$-field $\tilde B_{ij}$ of the dual background \eqref{eqn:tildeEij}. For the generalized dilaton $\widetilde{\hat d} = d + 1/2 \log v$ holds. Finally, assuming a constant fluctuation part of $\chi$, the  R/R field strengths spinor transforms covariantly 
\begin{equation}\label{eqn:trafoRR}
  \widetilde{\slashed{\partial}} \widetilde{\widehat{\chi}} = S_{\widehat{O}} \,      
    \slashed{\partial}{\widehat{\chi}}\,,
\end{equation}
like $\chi$. Clearly the two dual backgrounds share the same action on the Drinfeld double $\mathcal{D}$. Hence, the DFT version presented in this letter makes Poisson-Lie T-duality manifest like the conventional formulation does for abelian T-duality. 

In order to compare the R/R sector results with the literature, we take $\tilde G$ to be abelian. Now, $\widehat{O}$ admits the decomposition
\begin{equation}
  \widehat{O}^{\hat I}{}_{\hat J} = \frac12 \begin{pmatrix} 
      \tilde e^T & -\tilde B \tilde e^{-1} \\ 
      0 & \tilde e^{-1} \end{pmatrix}
  \begin{pmatrix} 
    1 - \Lambda & 1 + \Lambda \\
    1 + \Lambda & 1 - \Lambda \end{pmatrix}
  \begin{pmatrix}
    e^{-T} & 0 \\
    0 & e
  \end{pmatrix}{}^{\hat I}{}_{\hat J}
\end{equation}
where $\tilde e^T \tilde e = \tilde g$ and $e$ is the left-invariant Maurer-Cartan form on $G$. Note that the $\Lambda$ appearing here is the one found in (2.8) of \cite{Sfetsos:2010uq} by comparing the chiral and anti-chiral components of the world-sheet current \eqref{eqn:Ja}. Starting for this observation, it is straightforward to show that \eqref{eqn:trafoRR} matches their transformation prescription.

In summary it is evident that DFT on a Drinfeld double is a natural proposal for a low-energy effective target space theory with manifest Poisson-Lie (and therewith non-abelian) T-duality. It provides a powerful framework to address the questions outlined in the first paragraph of this letter. Merging these two lines of research seems now feasible and hopefully will produce many new insights on both sides.

\begin{acknowledgments}
\paragraph*{Acknowledgements:} I would like to thank the organizers and participants of the workshop ``Recent Advances in T/U-dualities and Generalized Geometries'' 6-9 June 2017, Zagreb, Croatia for the inspiring atmosphere and for highlighting the significance of Poisson-Lie and non-abelian T-duality to me. This work is supported by the NSF CAREER grant PHY-1452037 and NSF grant PHY-1620311.

\end{acknowledgments}

%Classification of Drinfeld doubles \cite{Hlavaty:2001fb,Snobl:2002kq}

\bibliographystyle{apsrev4-1}
\bibliography{literatur}

\end{document}